%% file: main.tex
\documentclass[acmsmall]{acmart}
\usepackage{bbding}
\usepackage{bbm}
\usepackage{xspace}
\usepackage[T1]{fontenc}
\usepackage{multirow}
\AtBeginDocument{%
  \providecommand\BibTeX{{%
    \normalfont B\kern-0.5em{\scshape i\kern-0.25em b}\kern-0.8em\TeX}}}

\setcopyright{acmcopyright}
\copyrightyear{2022}
\acmYear{2022}
\acmDOI{XXXXXXX.XXXXXXX}

\begin{document}
\title{Distributed Energy Management and Demand Response in Smart Grids: A Multi-Agent Deep Reinforcement Learning Framework}


\author{Amin Shojaeighadikolaei}
\email{amin.shojaei@ku.edu}
\orcid{0000-0003-3772-1904}
\affiliation{
  \institution{University of Kansas}
  \city{Lawrence}
  \state{KS}
  \country{USA}
  \postcode{66045}}
\author{Arman Ghasemi}
\email{arman.ghasemi@ku.edu}
\affiliation{
  \institution{University of Kansas}
  \city{Lawrence}
  \state{KS}
  \country{USA}
  \postcode{66045}}
\author{Kailani Jones}
\email{kailanij@ku.edu}
\affiliation{
  \institution{University of Kansas}
  \country{USA}}
\author{Yousif Dafalla}
\email{yousif.dafalla@ku.edu}
\affiliation{
  \institution{University of Kansas}
  \country{USA}}
\author{Alexandru G. Bardas}
\email{alexbardas@ku.edu}
\affiliation{
  \institution{University of Kansas}
  \country{USA}}
\author{Reza Ahmadi}
\email{reza.ahmadi.ee@outlook.com}
\affiliation{
  \institution{Amazon Web Services, Kirkland}
  \country{USA}
}
\author{Morteza Hashemi}
\email{mhashemi@ku.edu}
\affiliation{
  \institution{University of Kansas}
  \country{USA}
}
\email{mhashemi@ku.edu}

\renewcommand{\shortauthors}{Shojaeighadikolaei, et al.}

\input{chapters/1.Abstract}


\maketitle

\input{chapters/intro_new}

\input{chapters/RelatedWork_new}

\input{chapters/4.ProblemFormulation}

\input{chapters/5.RL}

\input{chapters/6.Numerical}

\input{chapters/7.Conclusion}

\bibliographystyle{ACM-Reference-Format}
\bibliography{sample-base}

\end{document}

%% file: chapters/1.Abstract.tex
\begin{abstract}
This paper presents a  multi-agent Deep Reinforcement Learning (DRL) framework for autonomous control and integration of renewable energy resources into smart power grid systems. In particular, the proposed framework jointly 
considers demand response (DR) and distributed energy management (DEM) for residential end-users.
DR has a widely recognized potential for improving power grid stability and reliability, while at the same time
reducing end-users' energy bills. However, the conventional DR techniques come with several shortcomings, 
such as the inability to handle operational uncertainties while incurring end-user disutility, 
which prevents widespread adoption in real-world applications. 
The proposed framework addresses these shortcomings by implementing DR and DEM based on real-time pricing strategy that is achieved using deep reinforcement learning.  
Furthermore, this framework enables the power grid service provider to leverage distributed energy resources (i.e., PV rooftop panels and battery storage) as dispatchable assets to support the smart grid during peak hours, thus achieving management  of distributed energy resources. Simulation results based on the Deep Q-Network (DQN) 
demonstrate significant improvements of the 24-hour accumulative profit for both prosumers and 
the power grid service provider, as well as major reductions in the utilization 
of the power grid reserve generators. 
\end{abstract}

%% file: chapters/intro_new.tex
\section{Introduction}
Recent proliferation in renewable energy sources (RESs) and distributed energy resources (DERs) along with the advances in information and communication technologies (ICT) has  facilitated a paradigm  shift  from  the traditional 
power consumers to the more resourceful energy \emph{prosumers}. 
A prosumer is an active end-user with the ability to consume and produce energy~\cite{5759167}. The number of prosumers is increasing as more households are proceeding to install  solar photovoltaic (PV) rooftop panels and energy storage system (ESS)~\cite{ 7820171}. The premise of such a distributed network of energy resources is to provide economic and ancillary benefits 
to both prosumers and power grid service provider (SP)~\cite{khoshjahan2021impacts}. The interaction between the prosumers and SP has been investigated in several prior studies. Many works utilize model-based control paradigms, such as model predictive control (MPC)~\cite{gan2020data}, mixed-integer linear programming (MILP)~\cite{7114268}, dynamic and stochastic programming~\cite{farzaneh2019robust}, and alternating direction method of multiplier (ADMM)~\cite{7501891}. 
One key step for applying these techniques is to establish an accurate model that captures the dynamics of the microgrid and the interactions between various components with all the operational constraints.


 The massive deployment of DERs and RESs dramatically increases the complexity and uncertainty of the system due to the more complex cross-area power balancing between SP and prosumers. This makes the accurate system model hard to obtain. The model-based nature of the aforementioned methods also exhibits limited generalization capabilities~\cite{dorokhova2021deep}. Difficulties in handling nonlinear behavior, volatile renewable generations, and heterogeneity of the end-users represent other major shortcomings of these methods. To overcome these challenges, model-free Reinforcement Learning (RL) techniques are proven beneficial for demand-side energy management since they do not require an explicit model of the environment. In general, RL frameworks are emerging as the pre-eminent tool for sequential decision-making problems within unknown environments. Similar to many other scientific and engineering domains, the RL-based solutions are receiving more attention from the power system society. For instance, voltage and frequency control \cite{wang2021multi}, market bidding \cite{pedasingu2020bidding}, microgrid energy management \cite{samadi2020decentralized}, and demand response (DR) \cite{vazquez2019reinforcement} are just a few examples of power-related problems that can be solved using RL.


In this paper, and to bridge the gap between the power system and RL communities, we investigate the mutual interplay between participants and the different tasks in a residential microgrid. We propose a framework based on Deep Reinforcement Learning (DRL) for both SP and prosumers to enable dynamic decision-making.  A service provider agent (SPA) is deployed for solving the distributed energy management (DEM) problem in the microgrid by dynamically determining  the electricity buy price as a control parameter for optimization of the energy management across distributed prosumers. For instance, a higher electricity buy price during peak hours (e.g., evening hours) incentivizes the prosumers with surplus energy to discharge their battery. As a result, the prosumer receives economic benefit in terms of electricity bill reduction. This process is called Demand Response (DR).  The purpose of DR is to encourage the prosumers to actively participate in a program and contribute to the optimal energy distribution in the electricity retail market.  Moreover, buying electricity from distributed prosumers would enable the power grid service provider to support higher demands during peak hours. In our envisioned system model, the prosumer agent (PA) solves the prosumer energy management problem by determining the charge/discharge of the battery installation. Thus, each prosumer independently  maximizes its own profit. 

In summary, our main contributions are as follows:
\vspace{-0.15cm}
\begin{itemize}
 \item We formally define the interaction between SP and the prosumers as a Markov Decision Process (MDP), and develop a multi-agent DRL framework that interweaves the real-time energy management over the microgrid with the prosumer side real-time DR, using demand-dependent dynamic pricing for incentivizing prosumers' participation in DR.
  \item We show that the proposed DRL framework enables the service provider to leverage energy storage as dispatchable assets, while incentivizing the prosumers to actively participate in the program to support the grid activities.
 \item Our numerical results based on Deep Q-Network (DQN) demonstrate that the proposed framework reduces the average daily bill for prosumers, while at the same time it provides higher profits for the grid service provider (SP) by leveraging distributed energy resources instead of tapping into traditional reserve generation facilities with higher costs.  
\end{itemize}
The remainder of this paper is organized as follows. Section~\ref{sec:related_work} 
covers related work, followed by the system model and problem formulation in Section~\ref{sec:model}. 
Next, Section~\ref{sec:RLmodel} formulates the DRL framework and Section~\ref{sec:result} presents numerical results. Finally, Section \ref{sec:conclusion} concludes the paper.




%% file: chapters/RelatedWork_new.tex
\section{Related work}
\label{sec:related_work}

The interaction between service provider and end-users have been investigated in many prior works. Recently, there has been growing interests in adopting RL to address the concerns of this interaction. These attempts can be categorized as follows:

\textbf{Prior Works Focused on Service Provider}: 
A multitude of prior works use RL in the context of power systems, mainly to address service provider concerns. 
For example, a hierarchical electricity market with bidding and pricing over wholesale and retailer using DRL was proposed 
in~\cite{xu2019deep}. In ~\cite{liang2020agent}, a Deep Deterministic Policy Gradient (DDPG) algorithm was used to solve the bidding problem of several generation companies. Moreover, RL has been widely employed for energy management optimization~\cite{foruzan2018reinforcement, dridi2021novel}. In~\cite{foruzan2018reinforcement}, the authors proposed a multi-agent distributed energy management using Q-learning framework with time-of-use pricing for a microgrid energy market with a centralized battery pack and renewable generation. The authors in \cite{dridi2021novel} designed a novel RL method that uses classical recurrent neural networks instead of SAR (State-Action-Reward) method to solve the microgrid energy management in the Industrial Internet of Things (IIoT).
In contrast to these works, we propose a dynamic pricing scheme with DQN at the service provider side, combined with distributed battery and PV installations at the demand side.

\textbf{Prior Works Focused on End-users}: 
The authors in \cite{wang2020deep} developed a DDQN-based load scheduling
algorithm with a time-of-use scheme to reduce the peak load demand and the operation cost for the distribution system, however without considering prosumers. In~\cite{chung2020distributed}, the interaction between households and the service provider was modeled to jointly solve the energy scheduling consumption and preserving privacy for the households by using policy gradient RL method.
Furthermore, \cite{9281326} developed a home energy management system (HEMS) using Actor-Critic technique based on adaptive dynamic programming.
Likewise, in \cite{7401112}, an energy management DR using Q-learning adjusts 
the consumption plan for thermostatically controlled loads. 
Along the same lines, \cite{xu2020multi} proposed a framework for home energy management based on multi-agent Q-learning to minimize the electricity bill as well as DR-induced dissatisfaction costs for end-users. 
Their method considered proliferation of rooftop PV systems, but did not consider home energy storage systems. Moreover, the works in \cite{sun2020continuous,yang2019large}, focused on reducing the high peak load in large-scale HEM using Entropy-based multi-agent deep reinforcement learning, and the work in \cite{ye2020model} applied prioritized DDPG in multi-carrier energy system to solve the real-time energy management problem.
It is also worth noting that these works \cite{wang2020deep,chung2020distributed,9281326,7401112,xu2020multi,sun2020continuous,yang2019large} do not consider the grid side and only propose optimization algorithms for end-users.

\textbf{Prior Works Focused on Both Service Provider and End-users}: Other research works have addressed the interactions between retailer/SP and end-users. 
For example, \cite{samadi2014real} investigated a gradient-based method to minimize the 
aggregate load demand by assuming the presence of consumption scheduling devices on the users side.   
Furthermore, works such as \cite{jia2016dynamic, yu2018incentive,8490840} leveraged the Stackelberg game
to model and maximize 
the retailer profit and minimize the payment bill of its customers. 
The authors included renewable energy generation on the retailer side, which is different from our model 
with distributed PV installations across prosumers. 
The studies in \cite{dehghanpour2016agent} and \cite{lu2019incentive} proposed a hierarchical agent-based framework 
to maximize both retailer and customers profits. 
A Q-learning DR is proposed in \cite{DynamicPrice} for the energy consumption scheduling problem that can work without prior information and leads to reduced system costs. Also, in \cite{11Dynamic} both service provider profit and consumers' cost are considered to find the optimal retail price. However, unlike our effort, the works presented in \cite{dehghanpour2016agent,lu2019incentive,DynamicPrice,11Dynamic} only consider regular electricity consumers, rather than prosumers with generation and storage capabilities. 
In contrast to these efforts, we model a microgrid that consists of both consumers and prosumers that are equipped with battery storage. As a result, the proposed multi-agent DRL framework achieves DR with dynamic pricing and distributed energy management for a microgrid with large-scale rooftop PV panels and energy storage system.
Table~\ref{tab:relatedwork} compares our work 
with the previous related works. 

\begin{table}
 		\caption{Taxonomy of Related Work on Management of Distributed Energy Resources. (\Checkmark : Considered, ---: Not Considered ) }
 	\resizebox{\columnwidth}{!}{
  	\begin{tabular}{|l|c|c|c|c|c|}
  	\hline
  	\textbf{Reference} & \textbf{Energy management } & \textbf{Retailer/end-user} & \textbf{Renewable penetration}
  	& \textbf{Energy storage} & \textbf{RL} \\
  	 & \textbf{DR} & \textbf{profit optimization}  & \textbf{consideration} & \textbf{consideration}  & \textbf{algorithm}\\
  	 \hline
  	\cite{dehghanpour2016agent},\cite{DynamicPrice},\cite{11Dynamic} & \Checkmark & \Checkmark & --- & --- &\Checkmark \\
  	\hline
  	\cite{yu2018incentive},\cite{8490840},\cite{chung2020distributed},\cite{lu2019incentive}& \Checkmark & \Checkmark  & --- & --- & ---\\ 
  	\cite{xu2020multi} & ---& \Checkmark& \Checkmark  & --- & \Checkmark \\
  	\hline
  	\cite{sun2020continuous}, \cite{yang2019large} & --- & \Checkmark & --- & --- & \Checkmark\\
  	\hline
  	\cite{jia2016dynamic} & \Checkmark & --- & --- & \Checkmark & \Checkmark \\
  	\hline
  	Our proposed framework & \Checkmark &  \Checkmark & \Checkmark & \Checkmark  & \Checkmark \\ 
  	\hline
 	\end{tabular}}
 	\vspace{-0.1in}
 	\label{tab:relatedwork}
 \end{table}

%% file: chapters/4.ProblemFormulation.tex
\section{Microgrid Model and Problem Formulation}\label{sec:model}

\begin{figure}[t]
\centering
\includegraphics[scale=0.5, trim=1.05cm 0.5cm 0.9cm 0.5cm, clip]{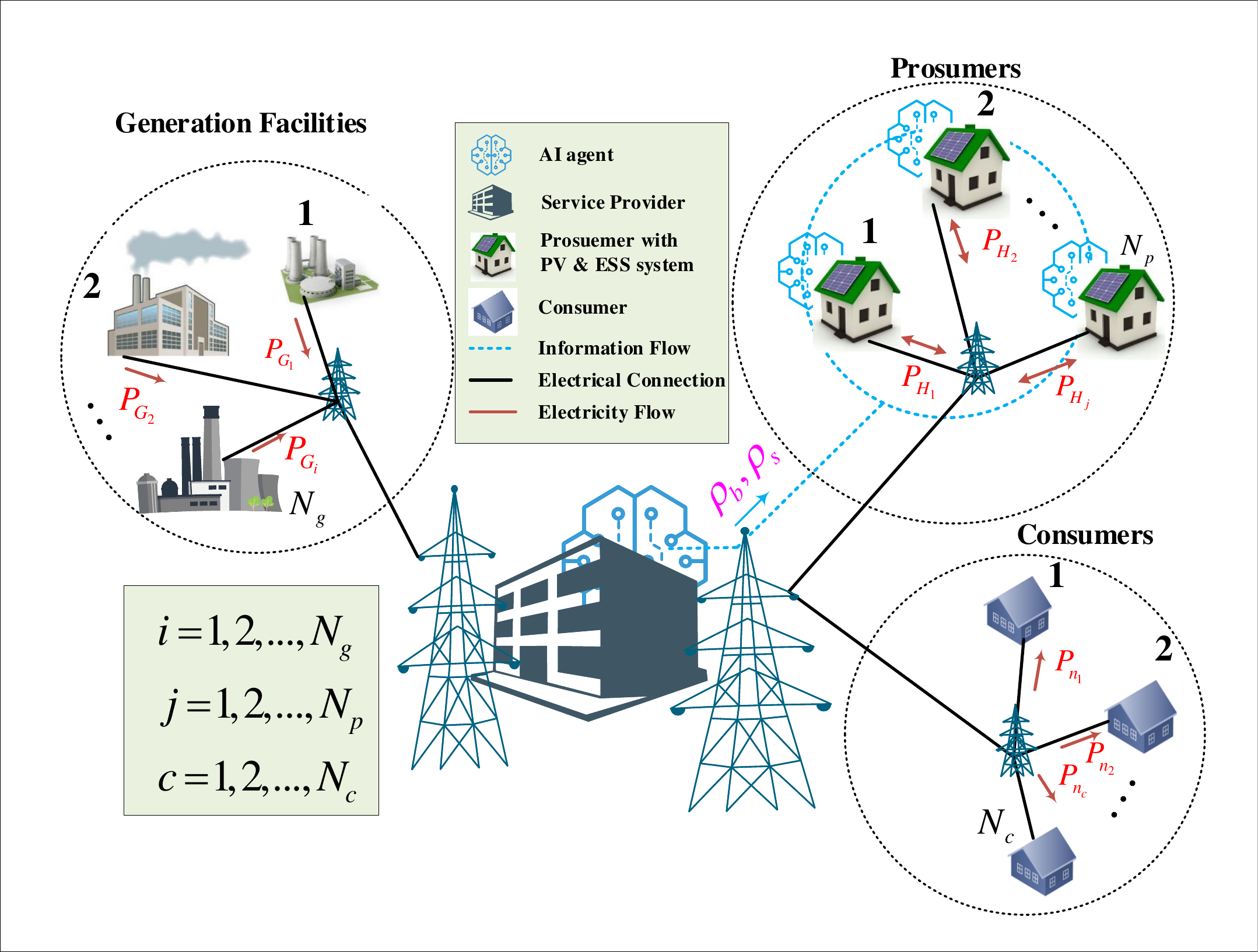}
\caption{Proposed microgrid system architecture that consists of generation facilities, traditional consumers, and prosumers that are equipped with solar rooftop panels and energy storage and deep reinforcement learning agents. Grid and prosumers are equipped with deep reinforcement learning agents to dynamically adjust their policies in terms of buy/sell prices and power injection.}
\vspace{-0.12in}
\label{fig:system-model}
\end{figure} 
The envisioned microgrid model is shown in  Figure~\ref{fig:system-model}. 
On the end-user side, the microgrid consists of regular consumers as well as prosumers that are equipped with PV panels, battery, and an intelligent agent.
The presence of consumers ensure that the microgrid always has energy deficiency. The conventional generation units are connected to the microgrid to meet the deficiency. The base generation units have low cost for the service provider but if their capacity is not enough, the reserve and the more expensive generation facilities will be dispatched to meet the real-time microgrid demand~\cite{goodarzi2022evaluate}. Without loss of generality, this paper forgoes the ramp rate constraints of the generation facilities.
The agent decides on whether to consume the excess energy generated by the PV system, store it in the battery, or sell it to the grid. On the other side, the SP is responsible 
for dispatching power to the loads from various generation facilities, including distributed energy resources of prosumers.

In this paper, distributed energy management and demand response optimization problems are defined for the SP and prosumers,  respectively. On the grid side, the SP is 
engaged in energy management with the objective of improving its own profit. On the household side, prosumers engage in DR 
with the goal of reducing their daily electricity bill. The complex interactions between the two sides 
denote the joint DEM-DR approach proposed in this paper. 
\textbf{Service Provider Energy Distribution Model:} In the presence of prosumers, the DEM problem of the microgrid evolves into a much more complex problem since the prosumers also appear as generation facilities in certain time periods due to injection of power into the grid. Hence, in this paper, the DEM formulation is defined as follows \vspace{-0.1in},
\begin{align}
   &\max_{\rho_b}
   \begin{aligned}[t]
      & \ \ {{{\Psi}}_G}(T)={{\mathop{\rm R}\nolimits} _G}(T) - \left\{ {\sum\limits_{i = 1}^{{N_g}} {{{\mathop{\rm F}\nolimits} _{{G_i}}}(T)}  + \sum\limits_{j = 1}^{{N_p}} {{{\mathop{\rm F}\nolimits} _{{H_j}}}(T)} } \right\} \label{eq:Grid Profit}
   \end{aligned} \\
   &\text{Subject to:} \notag \\
   & \sum\limits_{i=1}^{{N_g}} {{{\mathop{\rm F}\nolimits} _{{G_i}}}(T)}  = \sum\limits_{i=1}^{{N_i}} {{{\mathop{\rm F}\nolimits} _{{B_i}}}(T)}  + \sum\limits_{i=1}^{{N_r}} {{{\mathop{\rm F}\nolimits} _{{R_i}}}(T)} \label{eq:Reserve}, \\
   & {{\mathop{\rm R}\nolimits} _G}(T) = \int\limits_0^T {{P_D}(t).{\rho _s}(t)dt} \label{eq:Grid revenue}, \\
   & {{\mathop{\rm F}\nolimits} _{{H_j}}}(T) = \int\limits_0^T {{P_{{H_j}}}(t)} .{\rho _b}(t)dt\,\,\,\,\,\,\,\,\rm{for}\,\,\,{P_{{H_j}}}(t) \ge 0 \label{eq:Prosumer injection}, \\
   & {P_D}(t)  = \sum\limits_{i = 1}^{{N_g}} {{(P_{{G_i}}}(t)- P_{{loss}_i}(t))}  + \sum\limits_{j = 1}^{{N_p}} {{P_{{H_j}}}(t)} \label{eq:Power balance grid}, \\
   & P_{{loss}_i}= \beta_i \times P_{G_i}^2(t), \\
   & P_{{G_i}}^{\min } \le {P_{{G_i}}}(t) \le P_{{G_i}}^{\max }\,\,\,\,{\rm{for}}\,\,i = 1,2,...,{N_g} \label{eq:utility_limit},
 \end{align}
where ${{{\Psi}}_G}(T)$ denotes the SP profit over a time horizon of $T$, and ${{{\rm R}}_G}(T)$
is SP revenue as a result of selling electricity to the loads, which is calculated by Equation \eqref{eq:Grid revenue} where ${P_D}(t)$ and ${\rho_s}(t)$ denote the total system demand and electricity sell price at any given time, respectively. ${\rm F}_{G_i}(T)$ is the cost of generation unit $i$ defined as the quadratic function of $P_{{G_i}}$, which means ${\rm F}_{G_i}(P_{{G_i}})={a_i}P_{{G_i}}^2+{b_i}P_{{G_i}}+{c_i}$ where $a_i$, $b_i$, and $c_i$ are fitting parameters and $P_{{G_i}}$ is the amount of power purchased from generation unit $i$ \cite{ongsakul2019artificial}. This cost consists of the cost of base generation units ${{{\mathop{\rm F}\nolimits} _{{B_i}}}(T)}$ and cost of reserve generation units ${{{\mathop{\rm F}\nolimits} _{{R_i}}}(T)}$ in Equation \eqref{eq:Reserve} where $N_i$ and $N_r$ are the number of base and reserve generation facilities, respectively. $N_g$ is the number of all generation facilities. If the power demand exceeds the amount of base generation, the overall cost will be significantly higher because procuring reserve power is more costly. ${\rm F}_{H_j}{(T)}$ is the cost of buying electricity from the $j^{th}$ prosumer and calculated by Equation \eqref{eq:Prosumer injection} where ${P_{{H_j}}}(t)$ and ${\rho_b}(t) $ denote the power injected by the $j^{th}$ prosumer and electricity buy price, respectively. Equation \eqref{eq:Power balance grid} illustrates the total generation and total demand power balance requirement, which needs to be maintained at any given time slot $t$. $N_p$ and $N_c$ denote the number of prosumers and consumers, respectively. $P_{{loss}_i}$ denotes the losses induced by generation unit $i$, where $\beta_i$ is the loss coefficient \cite{7387777}. In addition, the power output of each generation unit must not exceed its operation limits, which are described in Equation \eqref{eq:utility_limit}.

The SP's goal is to maximize its profit. To do so, the SP dynamically changes the buy price to incentivize the prosumers to sell their excess energy to the grid, especially during the peak demand hours. Hence, the control variable in DEM problem is the buy price  $\rho_b(t)$.

\textbf{Prosumer Side Demand Response Model:}
The objective of the prosumers is to minimize their daily electricity bill by engaging in DR as a response to the dynamic buy price controlled by the SP. Therefore, the prosumer side profit maximization is formulated as follows\vspace{-0.1in},

\begin{align}
  &\max_{P_{b_j}}
  \begin{aligned}[t]
      & \ \ {{{\Psi }}_{{H_j}}}(T)={{\mathop{\rm F}\nolimits} _{{H_j}}}(T) + \boldsymbol{\mathbbm{1}}_{\{P_{H_j}(t) < 0\}} \times {{\mathop{\rm C}\nolimits} _{{H_j}}}(T) \label{eq:Prosumer profit}
  \end{aligned} \\
  &\text{Subject to:} \notag \\
  & {{\mathop{\rm C}\nolimits} _{{H_j}}}(T) = \int\limits_0^T {{P_{{H_j}}}(t)} .{\rho _s}(t)dt\,\,\,\,\,\,\,\,\rm{for}\,\,\,{P_{{H_j}}}(t) < 0 \label{eq:prosumer_cost},\\
  & {{\mathop{\rm F}\nolimits} _{{H_j}}}(T) = \int\limits_0^T {{P_{{H_j}}}(t)} .{\rho _b}(t)dt\,\,\,\,\,\,\,\,\,\,\rm{for}\,\,\,{P_{{H_j}}}(t) \ge 0, \nonumber \\
  & {P_{{H_j}}}(t) = {P_{P{V_j}}}(t) - {P_{b_j}}(t) - {P_{{C_j}}}(t) \label{eq:Power balance prosumer}, \\
  & SoC_{{b_j}}^{\min } \le \frac{1}{{{C_{{b_j}}}}}\int\limits_0^t {{P_{b_j}}(\tau )d\tau } + So{C_{{b_j}}}(0) \le SoC_{{b_j}}^{\max } \label{eq:batttery_limit}, \\
  & P_{b_j}^\text{max, dis} \le P_{b_j}(t)\le P_{b_j}^\text{max, charge} \label{eq:battery charge/discharge constraint}, \\
  & 0\le\ P_{PV_j}\left(t\right)\le\ P_{{PV}_j}^\text{max} \label{eq:PV constraint}, \\
  & \left|{P}_{H_j}(t)\right|\le\ P_{H_j}^\text{max} \label{eq:injection constraint}, \\
  & So{C_{{b_j}}}(0),\quad SoC_{{b_j}}^{\min },\quad P_{H_j}^\text{max}\geq0, \notag
\end{align}
where ${\Psi}_{H_j}{(T)}$ is the profit of the $j^{th}$ prosumer and  $C_{H_j}{(T)}$ is the cost of buying electricity from the grid over the time horizon $T$, which is described by Equation \eqref{eq:prosumer_cost}. ${{\mathop{\rm F}\nolimits} _{{H_j}}}(T)$ describes the prosumer $j$ revenue as a result of selling power to the grid and is calculated by Equation \eqref{eq:Prosumer injection}. Each prosumer has an internal household-scale power system with a power balance equation of its own that needs to be satisfied at all times. This is formulated by Equation \eqref{eq:Power balance prosumer} where $P_{PV_j}$, $P_{b_j}$ and $P_{C_j}$ are the amount of PV generation, battery charging/discharging power, and power consumption of the $j^{th}$ prosumer, respectively. According to Equation \eqref{eq:Power balance prosumer}, $P_{H_j}$ can take a positive or negative value depending on the amount of PV generation and battery charge or discharge action. A positive (negative) value means the prosumer $j$ injects (purchases) power to (from) the grid. Hence, $\boldsymbol{\mathbbm{1}}_{\{P_{H_j}(t)<0\}}$ is one/zero when the injected power from prosumer $j$ is negative/positive. The state of charge of each prosumer battery should be maintained within a safe operational range as illustrated in Equation \eqref{eq:batttery_limit} where $C_{b_j}$ is the nominal battery capacity and ${SoC}_{b_j}\left(0\right)$ represents the initial state of charge of the battery. ${SoC}_{b_j}^{\min}$ and ${SoC}_{b_j}^{\max}$ are the lowest and highest allowable state of charge for the battery, respectively. Furthermore, the charging and discharging power of the battery is limited as Equation \eqref{eq:battery charge/discharge constraint}, where $P_{b_j}^\text{max, charge}$ and $P_{b_j}^\text{max, dis}$ are the maximum charge/discharge power ratings of the battery. The constraint in Equation \eqref{eq:PV constraint} is imposed by the physical limitations of the size of the PV system and shows that the amount of PV generation is limited by the $P_{{PV}_j}^\text{max}$. Dynamic power injection into the grid might cause instability in some cases. Thus, the maximum allowable power injection is limited by Equation \eqref{eq:injection constraint}.

\begin{figure}[t]
\centering
\includegraphics[scale=0.42, trim=5.5cm 5cm 5.1cm 5cm, clip]{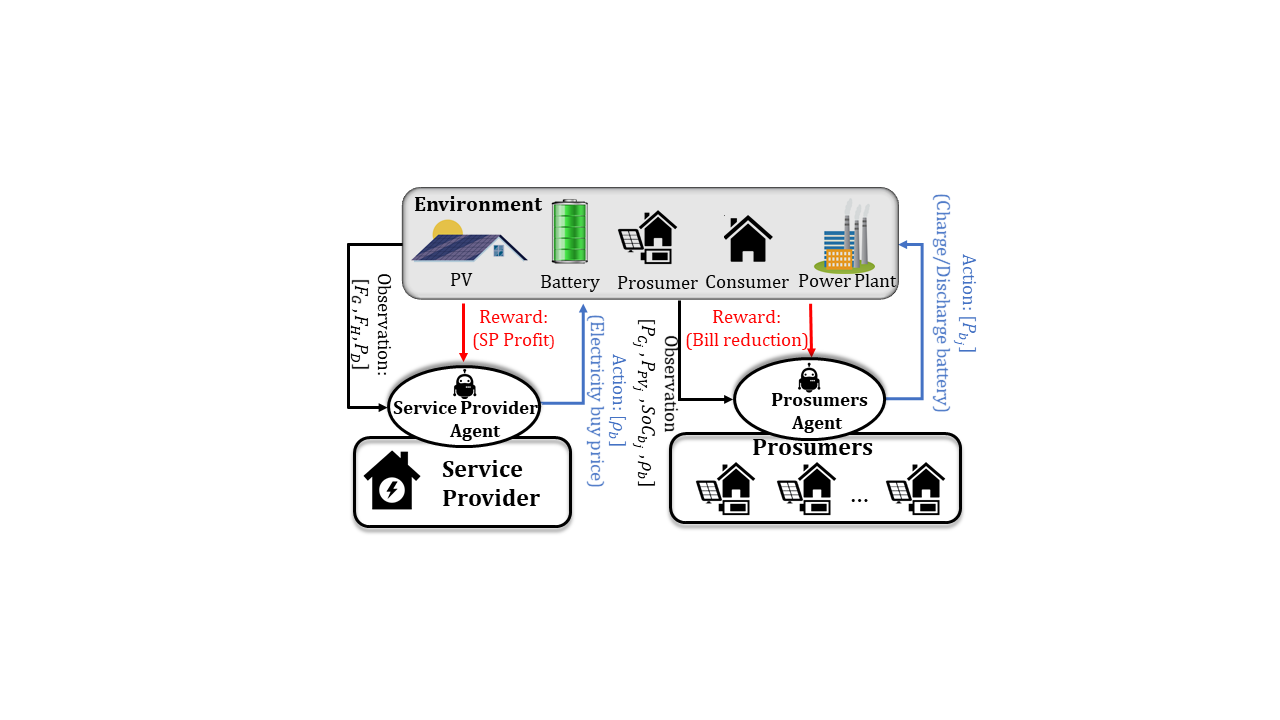}
\caption{Implemented Reinforcement Learning framework}
\label{fig:RL}
\end{figure}

In this paper, the formulated DR method is different from the conventional time-based DR, such as Real-Time Pricing (RTP) methods \cite{samadi2014real}.
Conventionally, RTP approaches rely on dynamically changing the electricity sell price to motivate the customers to alter their energy use profile, while the proposed DEM-DR algorithm relies on dynamically changing the buy price instead of the sell price. This potentially incentivizes prosumers to participate in the program without any customer dissatisfaction. At any given time slot $t$, both aforementioned optimization problems must be solved at the same time. Dynamic SP pricing scheme in DEM-DR problem reinforces the need to implement the optimization problems in short intervals (e.g., 15 minutes).

%% file: chapters/5.RL.tex
\section{Reinforcement Learning for DEM and DR}\label{sec:RLmodel}
Reinforcement Learning (RL) is a model-free machine learning technique that trains an agent to take optimal actions through repeated interactions with an environment.
The general RL framework consists of a set of states $\mathcal{S}$, a set of actions $\mathcal{A}$, a reward function $R$, and probability of transition between the states. The policy $\pi$ is a mapping from the states of the environment ($s^t$) to the actions ($a^t$), i.e., $\pi:\mathcal{S}\rightarrow\mathcal{A} $. 
The agent's ultimate goal is to learn which action $a^t \in \mathcal{A}$ to take at each time instance $t$ to maximize its
cumulative reward over time.  The action $a^t$ results in reward $r^t$, and the environment transitions from the state $s^t$ to $s^{t+1}\in \mathcal{S}$.

In this section, we use $t$ as superscript to denote discrete time slots. It should be noted that there is no interaction between the prosumers, such that their goals are to maximize their local profit regardless of other prosumers' performance.


\vspace{-0.1in}
\subsection{Decision-Making Problem Formulation} 
In order to tackle the optimization problems in Section \ref{sec:model}, we leverage a multi-agent DRL approach such that we define agent for the grid side SP and agents for the prosumer side with PV and battery storage capabilities. The $\text{SPA}$ is representative of the service provider agent, while $\text{PA}_j$ denotes the $j^{th}$ prosumer agent. Each agent observes the environment and subsequently takes an action, receiving a reward commensurate with the merit of the action. The agents will receive a reward through their action selection, which is profit maximization for SPA and electricity bill reduction for PA. Figure \ref{fig:RL} graphically demonstrates the interaction between the SPA and PA agent with the environment, their observation, and the action for each agent. 
\subsubsection{Service Provider Agent} 
The $\text{SPA}$ observes the cost of buying electricity (base and reserve generation) from $N_g$ generation facilities at time slot $t$, which is  denoted by ${\bf{F}}_G^t = [{\mathop{\rm F}\nolimits} _{{G_1}}^t,{\mathop{\rm F}\nolimits} _{{G_1}}^t,...,{\mathop{\rm F}\nolimits} _{{G_{{N_g}}}}^t]$. In addition, the $\text{SPA}$ observes the cost of buying electricity from $N_p$ prosumers denoted by ${\bf{F}}_H^t = [{\mathop{\rm F}\nolimits} _{{H_1}}^t,{\mathop{\rm F}\nolimits} _{{H_1}}^t,...,{\mathop{\rm F}\nolimits} _{{H_{{N_p}}}}^t]$, and the total power demand of the loads $P_D^{t}$ at the time slot $t$. Hence, environment states $s_{SPA}^t = \{ {\bf{F}}_G^t ,{\bf{F}}_H^t ,P_D^{t} \} \in \mathcal{S}_{SPA}$ are observable by the SPA. Subsequently, the SPA takes an action by adjusting the electricity buy price denoted by $a_{SPA}^t=\rho_b^t \in \mathcal{A}_{SPA}$, where $\mathcal{A}_{SPA}$ is the finite set of available actions for SPA, i.e., all the possible buy prices. 

The $\text{SPA}$ reward function is defined based on the grid profit, 
\vspace{-0.05in}
\begin{equation}\label{eq:SPA_reward}
r_{SPA}^t = P_D^t \times \rho _s^t - \sum\limits_{i = 1}^{{N_g}} {{\mathop{\rm F}\nolimits} _{{G_i}}^t - \sum\limits_{j = 1}^{{N_p}} {\boldsymbol{\mathbbm{1}}_{\{P_{H_j}(t)\geq0\}} \times P_{{H_j}}^t \times \rho _b^t}},
\end{equation}
where $r_{PSA}^t$ is the SPA reward at time slot t.
The value of ${\mathop{\rm F}\nolimits} _{{G_i}}^t$ is obtained using the incremental cost curve of the $i^{th}$ generation facility. Given the definition of the immediate reward, the ultimate goal for the SPA is to maximize the cumulative reward $R_{SPA}^t=\sum_{k=0}^{\infty}{\gamma^k r_\text{SPA}^{t+k+1}}$ over an infinite time horizon known as expected return,
where $0\le{\gamma}\le1$ is the discount factor.

\subsubsection{Prosumer Agent}  
${PA}_j$ observes its own local parameters at each time slot $t$ including power consumption $P_{C_j}^t$, state of charge of the battery $SoC_j^t$, and the PV generation $P_{PV_j}^t$, as well as the electricity buy price $\rho_b^t$ which is the result of SPA action. Hence $s_{PA_j}^t = \{ P_{C_j}^t,SoC_j^t ,P_{PV_j}^t , \rho_b^t \} \in \mathcal{S}_{{PA}_j}$.
Subsequently, the charge/discharge command to the energy storage in prosumer $j$ is the action determined by $PA_j$, which is shown by $a_{PA_j}^t=P_{b_j}^t\in\mathcal{A}_{PA_j}$. 
In this case, $\mathcal{A}_{PA_j}$ is the finite set of available actions to $PA_j$.
The reward function for the prosumer agents is defined as,
\begin{equation}\label{eq:PA reward}
r_{P{A_j}}^t = \boldsymbol{\mathbbm{1}}_{\{P_{H_j}(t)\geq0\}}  \times P_{{H_j}}^t \times \rho _b^t + \boldsymbol{\mathbbm{1}}_{\{P_{H_j}(t) < 0\}} \times P_{{H_j}}^t \times \rho _s^t.
\end{equation}

Similar to the SPA, the $PA_j$ tries to maximize its infinite-horizon accumulative reward $R_{{PA_j}}^t=\sum_{k=0}^{\infty}{{\tilde{\gamma}}_j^k r_{{PA_j}}^{t+k+1}}$,
where $0\le\tilde{\gamma}_j\le1$ is the discount rate for $PA_j$.

\begin{figure}[t]
\centering
 \includegraphics[scale=0.34, trim=4.5cm 1cm 5cm 0.04cm, clip]{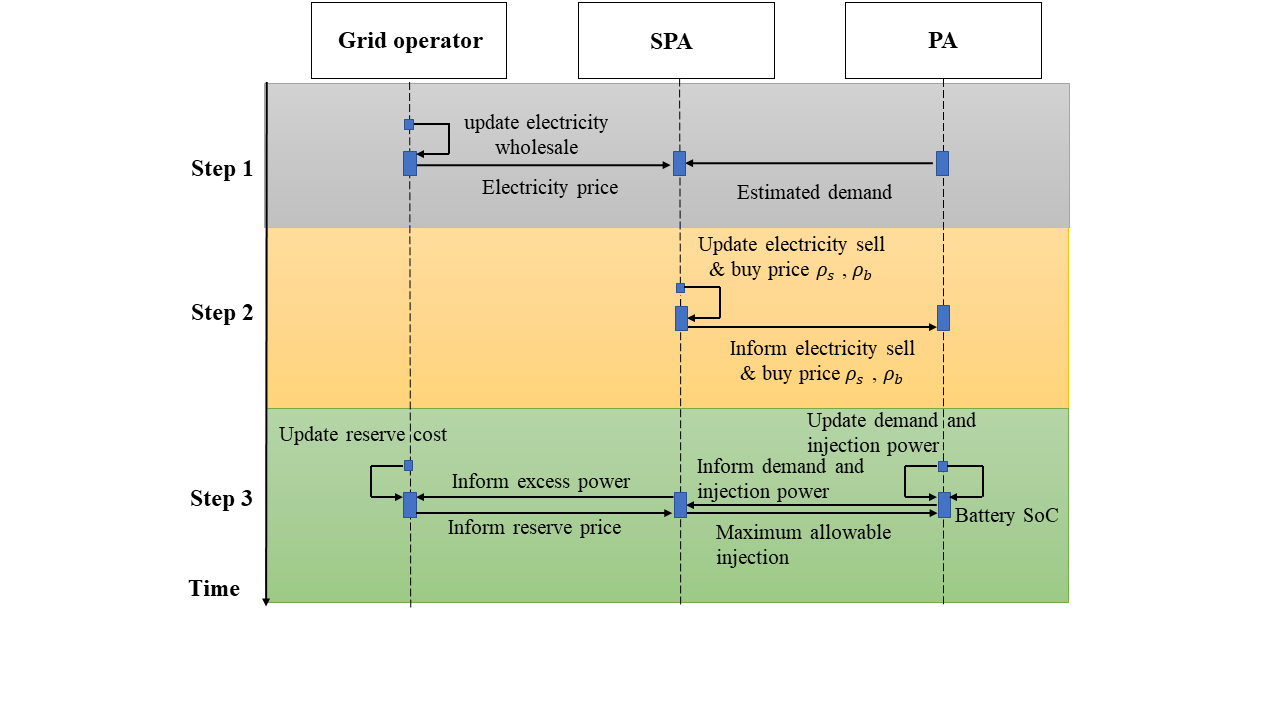}
 \vspace{-0.3in}
 \caption{Timeline of interaction between the SPA and PA. At each time slot $t$, the SPA dynamically determines the buy price in Step 2 (based on collected information in Step 1), and the PA decides on injected power and changing the battery state of charge in Step 3. }
 \vspace{-0.1in}
 \label{Timelinel}
\end{figure}


The timeline of interactions between the SPA and PA agents is illustrated in Figure~\ref{Timelinel}. At a given time slot $t$, the SPA makes a decision first by determining the market electricity buy price ($a_{SPA}^t$) based on the complete information of the prices of electricity generation facilities and the electricity demand from the prosumers' agents. The SPA then receives a reward ($r_{SPA}^t$) for taking the action. Next, the PAs observe the sell and buy price, and make the decision ($a_{PA_j}^t$) in terms of changing the battery state of charge and receive a reward ($r_{P{A_j}}^t$) for that decision. Finally, the PAs inform the SPA about the demand and the injected power.  


\vspace{-.1cm}
\subsection{Deep Q-Network Learning}
A systematic review of Deep Learning (DL) methods applied to the smart grid is provided in \cite{massaoudi2021deep}. Deep Reinforcement Learning is the combination of Deep Neural Networks (DNNs) and RL. In general, RL solutions can be divided into value- and policy-based algorithms. 
The value-based solutions are straight-forward to implement. 
The main value-based method is Q-learning. Due to the high dimensionality nature of the problem Deep Q-Network (DQN) agents, which can handle a large state space, are used to solve their respective Markov Decision Processes (MDPs) and maximize their accumulative rewards. In addition, since the action spaces of SPA and PAs are discrete, the DQN can be used for solving this problem. The transition formula of Q-learning is,
\begin{align}\label{eq:Q value}
Q(s^t,a^t) \leftarrow  Q(s^t,a^t) + {\alpha}[r^{t + 1} + {\gamma}&\mathop{\max }\limits_{{a ^{t + 1}}} Q(s^{t + 1},a^{t + 1}) - Q(s^t,a^t)]\ ,
\end{align}
where $\alpha$ is the learning rate. Additionally, $\gamma$ is the discount factor, with a range of 0 to 1. If $\gamma$ is closer to zero, the agent prioritizes the immediate reward over long-term rewards. On the other hand, if $\gamma$ is closer to one, the agent places greater weights over the future rewards compared with the immediate reward. The estimated Q-values are used to find the optimal policy that maximizes the accumulative rewards. 
An $\epsilon$-greedy strategy is used to balance between exploration and exploitation, i.e., 
\begin{equation*}\label{eq:E greedy}
    {a^t}=\begin{cases}
        \arg \mathop {\max }\limits_{{a^t}} \,E\left[ {Q\left( {{s^t},{a^t}} \right)} \right] & \text{with probability} \ 1 - \epsilon, \\
    \text{random action} & \text{with probability} \  \epsilon, 
            \end{cases}
\end{equation*}
where the probability of random action is $\epsilon$ that decays from 1 to 0.01 over the training episodes in our simulations. 

%% file: chapters/6.Numerical.tex
\section{Numerical Results}\label{sec:result}
In this study, we leverage the grid and prosumers' agents to implement the proposed DEM-DR scheme referred to as \emph{Agent-Based} scheme. To demonstrate the efficacy of the proposed approach, we compare it with the \emph{Conventional} approach, which simply injects the excess power to the grid when the battery is fully charged.  The simulation parameters for DQN agents are tabulated in Table~\ref{tab:SimulationParam}. As previously mentioned, the set of buy prices is defined as $\mathcal{A}_{SPA}$, representing the action space of the service provider agent. We designate the action space for prosumers as follows, 
\begin{align}\label{eq:PAaction}
P_{{b_j}}^t = \begin{cases}
                P_{b_j}^{\text{max,charge} } &\text{Charge\ action,} \\
                0 &\text{No\ charge\ or\ discharge\ action,}\\
                 P_{{b_j}}^{\text{max,dis} }&\text{Discharge\ action.}
            \end{cases}
\end{align}



\begin{table}[t]
\caption{DQN Hyperparameters and Simulation Parameters.}
\resizebox{0.7\columnwidth}{!}{
\centering
  {
    \begin{tabular}{l|l|l}
      \hline
      \textbf{Hyperparameters} & \textbf{Value for $SPA$} & \textbf{Value for $PA_j$}\\
      \hline
      Batch size & 64 & 64 \\
      Discount factor & ${\gamma}$=[0.95-0.99] & $\tilde{\gamma}_j$=[0.95-0.99] \\
      Learning rate & $\alpha$=1e-3 & ${\tilde{\alpha}_j}$=1e-3 \\
      Soft update interpolation & 1e-5 & 1e-5 \\
      Hidden Layer-nodes & 1-[1000] & 2-[1000,1000] \\
      Activation & Tanh & Tanh \\
      Optimizer & Adam & Adam \\
      \hline
      \hline
      \hline
      \textbf{Simulation Parameter} & \textbf{Description} & \textbf{Value}\\
      \hline
      $P_{PV_j}^{\max }$ & Max. PV Generation & [2-6] kW \\
      $P_{b_j}^{\text{max,charge} }/P_{b_j}^{\text{max,dis} }$ & Max. allowable charge/discharge & 2/-2.5 kW \\
      $P_{H_j}^{\max }$ & Max. allowable power injection & 10 kW \\
      ${SoC}_{b_j}^{\max }$ & Max. state of charge & $0.9 \times {C_{b_j}}$ \\
      ${SoC}_{b_j}^{\min }$ & Min. state of charge & $0.1 \times {C_{b_j}}$ \\
      $C_{b_j}$ & Energy storage capacity & [8-15] kWh \\
      ${\phi _j}(0)$ & Initial state of charge & [1-4] kWh \\
      $\rho _s$ & Sell price [before 11am, after 11am] & [0.05, 0.095] \$/kWh \\
      $\mathcal{A}_{SPA}$ & Buy price set for Agent-Based scenario & ${\{0.05,0.06,0.07,}$ \\ 
                  &                                    & ${0.08,0.09,0.1}\}$\$/kWh \\
      $\rho _b^t$ & Buy price for Conventional scenario & 0.06 \$/kWh \\
      $\left[ {P_{{G_1}}^{\min },P_{{G_1}}^{\max }} \right]$ & Limitation of base generation & [5, 45] kW  \\
      $\left[ {P_{{G_2}}^{\min },P_{{G_2}}^{\max }} \right]$ & Limitation of reserve generation & [0, 100] kW  \\
      $\left[ {\beta_1,\beta_2} \right]$ & Transmission loss coefficient of two generators & [0.0002, 0.0002] \\
      \hline
    \end{tabular}
    }}
 \label{tab:SimulationParam}
  \vspace{-0.1in}
\end{table}

\begin{figure}[t]
\centering
\includegraphics[scale=0.30, trim=0.2cm 0.2cm 0.2cm 0.2cm, clip]{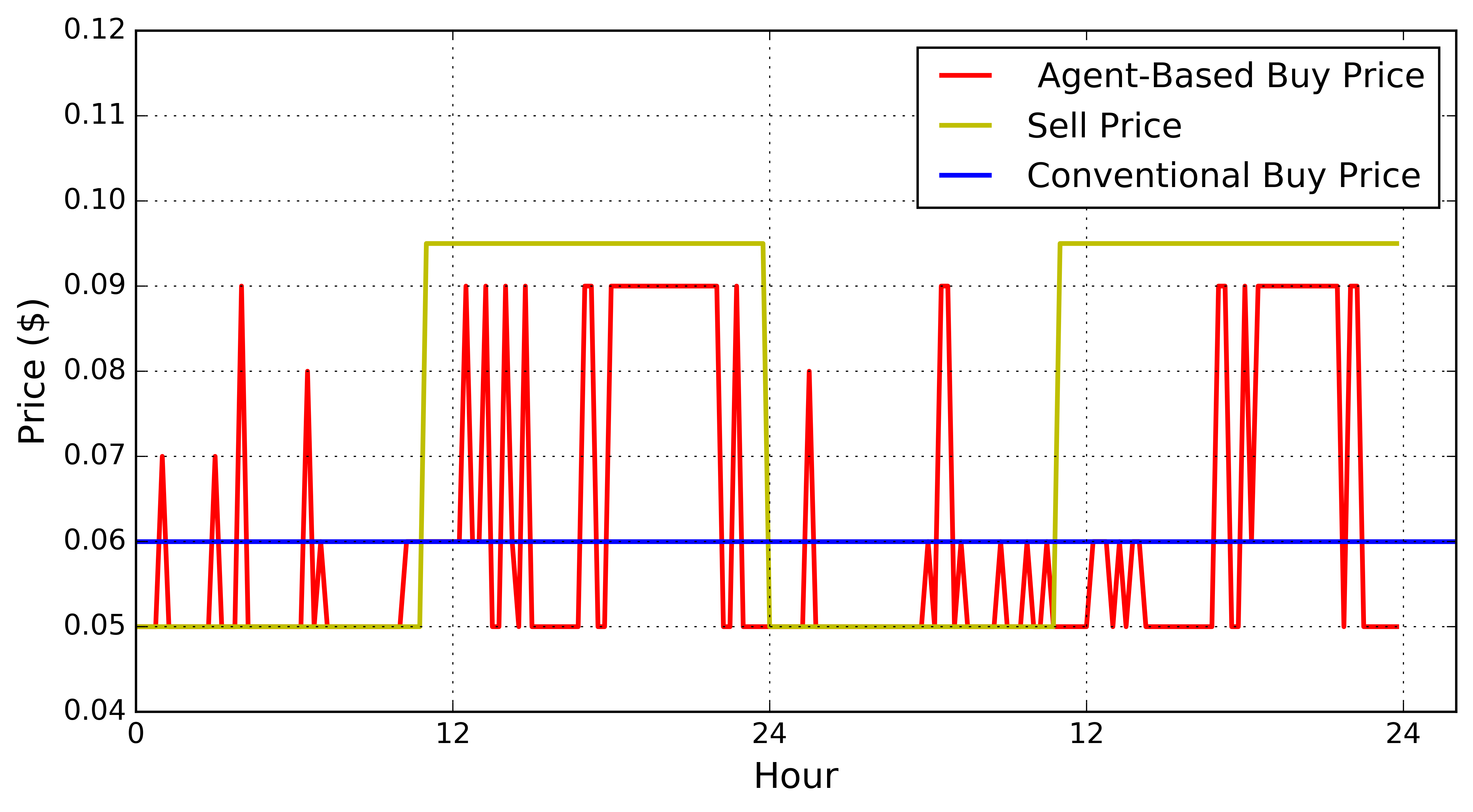}
\vspace{-.2cm}
\caption{Retail prices for the last two days of the simulation.}
\label{Price}
\end{figure}

\begin{figure}[t]
\centering
\includegraphics[scale=0.24, trim=3cm 2cm 0.9cm 2cm, clip]{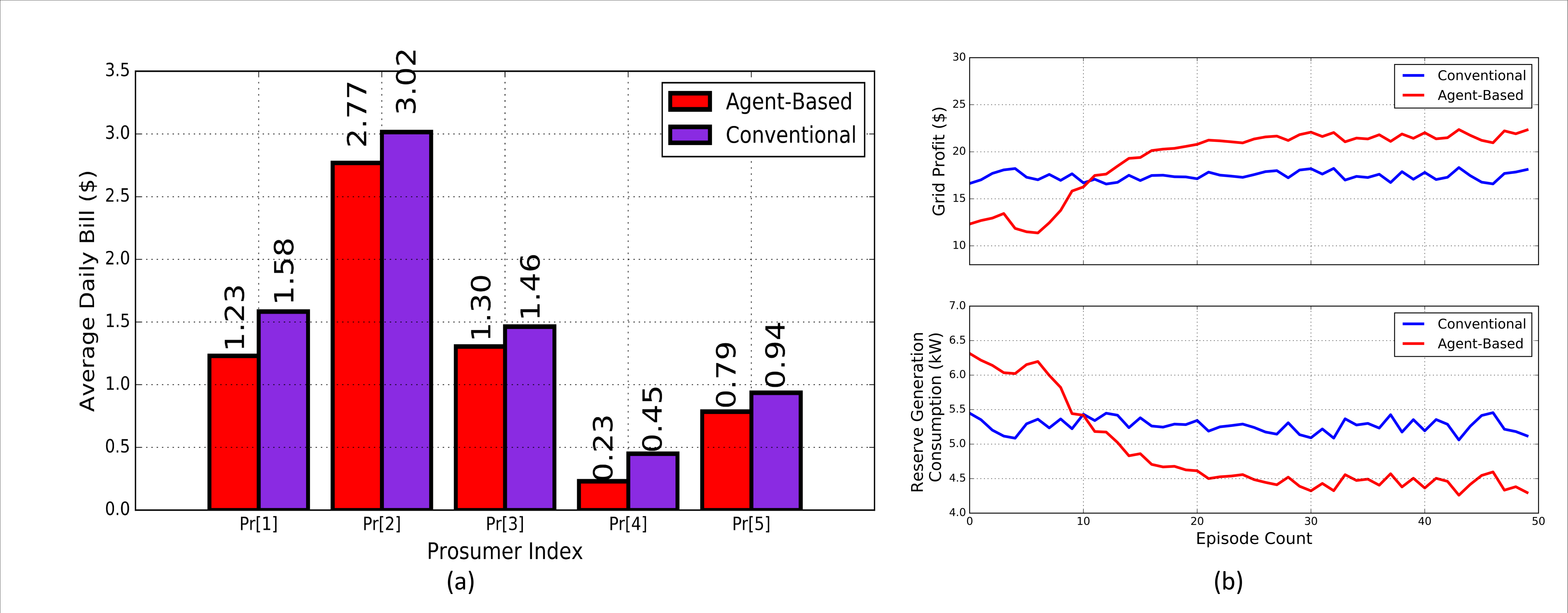}
\vspace{-0.1cm}
\caption{(a) Average daily bill comparison of the Conventional vs. Agent-Based schemes in a small-scale microgrid with 5 prosumers. (b) Performance comparison of the Conventional vs. Agent-Based scenarios in terms of grid profit and reserve unit utilization. }
\label{5prosumers_profit}
\vspace{-0.05in}
\end{figure}


We compare the Agent-Based with the Conventional scheme through two scenarios: \textbf{(1)} A small-scale setting with 5 prosumers in which we demonstrate the agents' effectiveness in interacting with the energy marketplace environment, and improvements made to both SP and prosumers' economic benefits. \textbf{(2)} A medium-scale setting with 50 prosumers wherein we investigate the system's reliability and scalability by simulating a larger-scale microgrid. We implemented our DQN agents using Python3 with PyTorch 1.5.0.
All simulations were performed via episodic updating across 10,000 episodes, each of which represents a 24-h cycle. A cycle consists of 96 iterations (sampling time is 15 minutes).
\vspace{-0.1in}
\subsection{Scenario I: Small-Scale Simulation with 5 Prosumers}
The first scenario studies the microgrid in Figure~\ref{fig:system-model} consisting of two generation facilities (a base plant and a spinning reserve) managed by one distribution SP, five prosumers, and one non-generational consumer. The DQN agents are implemented for the SP and the prosumers. The operational parameters of the five prosumers and the single consumer are tabulated in Table~\ref{tab:SimulationParam}. The utilized consumption and generation profiles for the prosumers mimic real-world  trends reported by California ISO \cite{CaliIso} to exemplify real-world operation.
\vspace{0.1in}
\subsubsection{\textbf{Macroscopic Evaluation of Prosumers and Grid Benefits}}
As reported in Table~\ref{tab:SimulationParam}, the sell price for both Conventional and Agent-Based scenarios takes two values depending on the time of a day. On the other hand, the buy price is constant during a day for the Conventional scenario, while in the Agent-Based scenario the buy price is dynamically determined by the SPA. Figure~\ref{Price} compares the dynamic pricing achieved by the Agent-Based approach vs. the fixed-pricing of the Conventional scheme. The results shown in Figure~\ref{5prosumers_profit} (a) compares the average daily bill of prosumers when utilizing the two approaches.  As pictured, the average daily bills have decreased by $22.1\%$, $8.2\%$, $10.9\%$, $48.8\%$, and $15.9\%$ for prosumer 1 to prosumer 5, respectively. Figure \ref{5prosumers_profit} (b) on the other hand, illustrates the moving average of SP profit and reserve generation consumption. According to these results, the Agent-Based approach is offering an $25.7\%$ increase in the SP profit, made possible by a $16\%$ reduction in reserve power consumption, compared to the conventional approach. This further demonstrates that the proposed Agent-Based approach is capable of providing a higher profit for the SP while offering greater economic benefit for the prosumers, warranting a win-win scenario for the grid and prosumers.  
\subsubsection{\textbf{Microscopic Evaluation of  Prosumers' Behavior}}
\begin{figure}[t]
\centering
\includegraphics[scale=0.39, trim=0.25cm 0.25cm 0.25cm 0.25cm, clip]{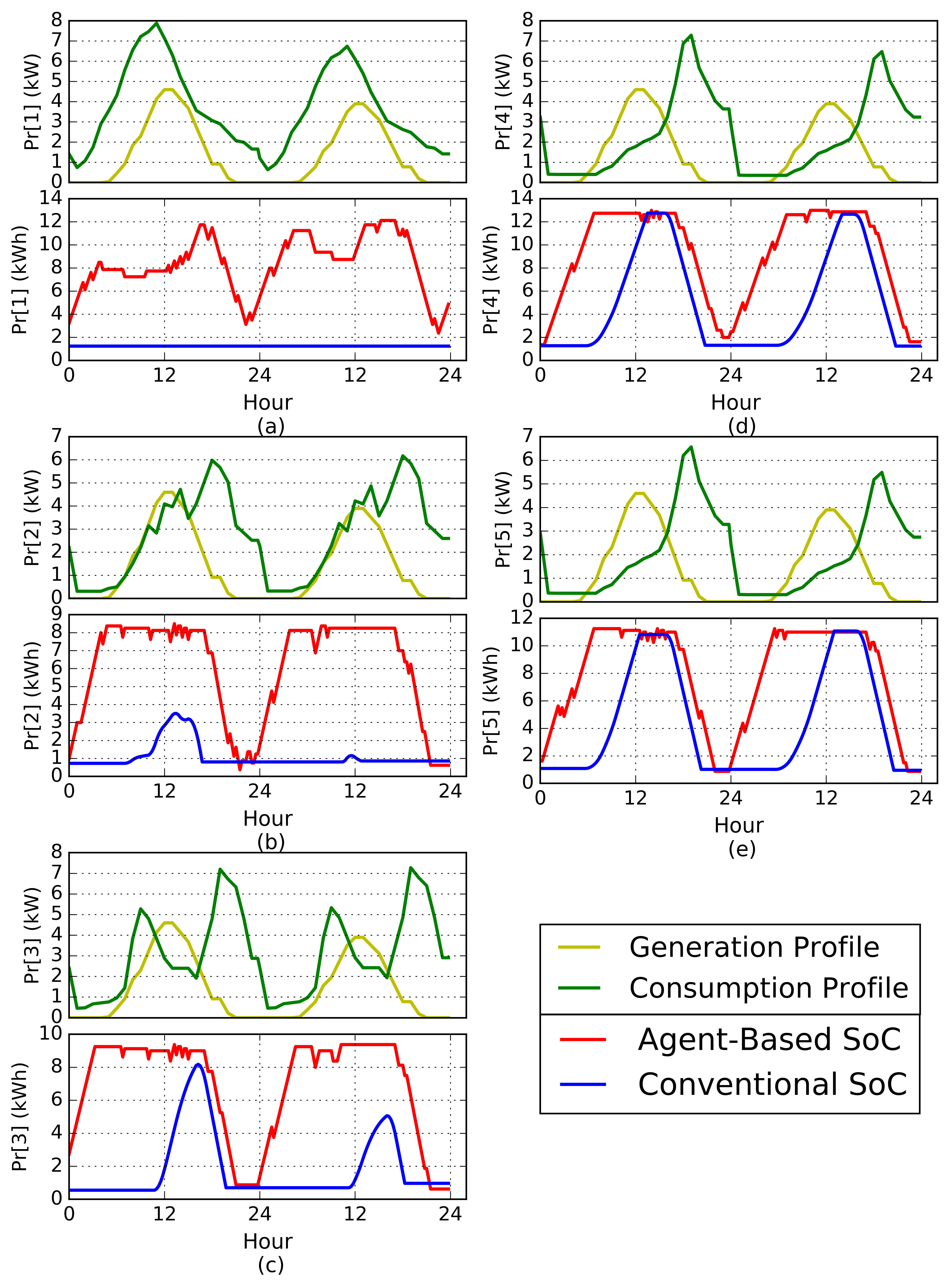}
 \vspace{-0.3cm}
\caption{Generation and consumption profiles along with the State-of-Charge (SoC) for 5 prosumers.}
\label{BattComparison}
 \vspace{-0.05cm}
\end{figure}

Figure~\ref{BattComparison} illustrates the temporal profiles of several internal states of all five prosumers over the course of the last two simulated days for the Agent-Based and Conventional approaches. These graphs provide an intuitive interpretation of how each demand-side participators respond to the DR scheme in real-time. As pictured, using the Conventional method, the households' PV systems can charge their batteries only when their generation is more than consumption, while only being able to sell their excess generation when their batteries are fully charged. 

\noindent \textbf{Observations on prosumers 1 and 2:} As demonstrated in Figure~\ref{BattComparison} (a) and (b), the generated power by prosumers 1 and 2 is less than their consumption (except for a very short period of time for prosumer 2). Therefore, throughout simulation of the Conventional approach, their batteries are never charging, denying them potential economic benefits otherwise possible by incorporating the energy storage system, and prohibiting them from participating in grid support during peak demand hours. On the other hand, using the Agent-Based approach, the prosumer batteries are charging during off-peak-low-price hours (i.e., beginning of the day), and prosumers are engaging in grid support during peak demand hours.  Although charging the battery at the beginning of the day incurs some cost for prosumers, it eventually provides higher economic benefits to them through selling the energy back to the grid at a higher price, while assisting with grid power balance during peak demand hours. 
In other words, the stored energy in prosumer 1 and 2 batteries is dispatched to the grid by properly incentivizing them to participate in DR.
 
\noindent \textbf{Observations on prosumer 3:} According to consumption and generation profiles for prosumer 3 illustrated in Figure~\ref{BattComparison} (c), only a small amount of excess PV generation during a few hours is discernible. In this case, the results demonstrate that deploying the Agent-Based algorithm yields a higher SoC compared with the Conventional method. 
This is because the RL agent learns to charge the battery at the beginning of the day when the selling price by the grid is low, and to support the grid in the afternoon when the buy price is high. 

\noindent \textbf{Observations on prosumers 4 and 5:} According to Figure~\ref{BattComparison} (d) and (e), these prosumers have excess PV generation during peak sun hours (e.g., around noon). The Conventional scheme fully charges the batteries of these prosumers during the peak sun hours, while the Agent-Based scheme charges the batteries starting from the beginning of the day when the sell price is low. As a result, these prosumers can sell their excess power during sun hours, rather than storing it, at even a higher price rate than the Conventional scheme could, after fully charging the batteries. 




\vspace{0.1in}
\subsubsection{\textbf{Effect of Battery Size}}
\begin{figure}[t]
\centering
\includegraphics[scale=0.29, trim=1.2cm 0.2cm 0.5cm 2cm, clip]{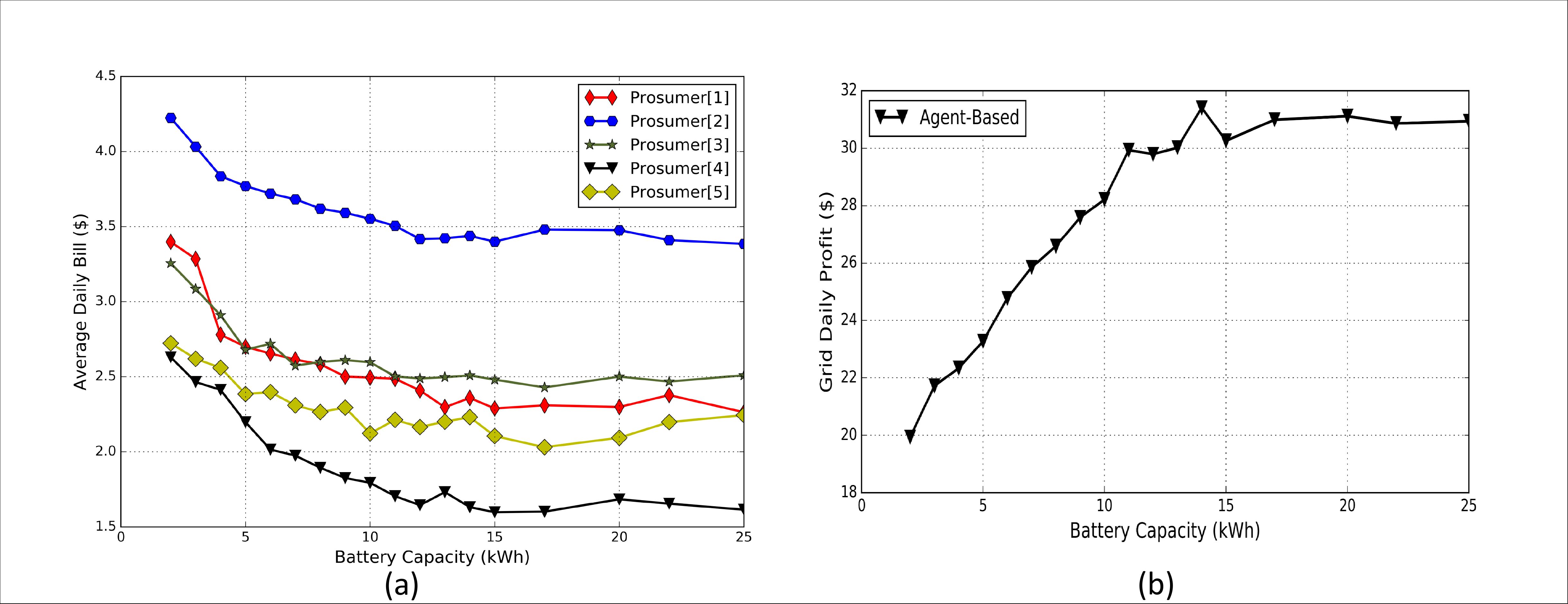}
\vspace{-.5cm}
\caption{(a) Effect of increasing battery capacity on the prosumers' average daily bill reduction in small-scale simulation. (b) Effect of increasing battery capacity on the grid profit improvement in small-scale simulation. 
}
\label{Battsize}
\end{figure}

 Figure~\ref{Battsize} demonstrates the prosumers' daily bill reduction and grid daily profit achieved by increasing the size of prosumer batteries from $2$kWh to $25$kWh.  As pictured, the  daily bill for all five prosumers decreases as the battery capacities are increased. Similarly, the grid daily profit also increases by increasing the battery capacities. Nevertheless, the improvements start to level down as the battery capacities approach to around $15$kWh. Based on this observation, we use batteries with capacities in the range of $8$kWh to $15$kWh for all prosumers in the simulations. 

\begin{figure}[t]
\centering
\vspace{-.3cm}
\includegraphics[scale=0.31, trim=1cm 0.25cm 0.5cm 1cm, clip]{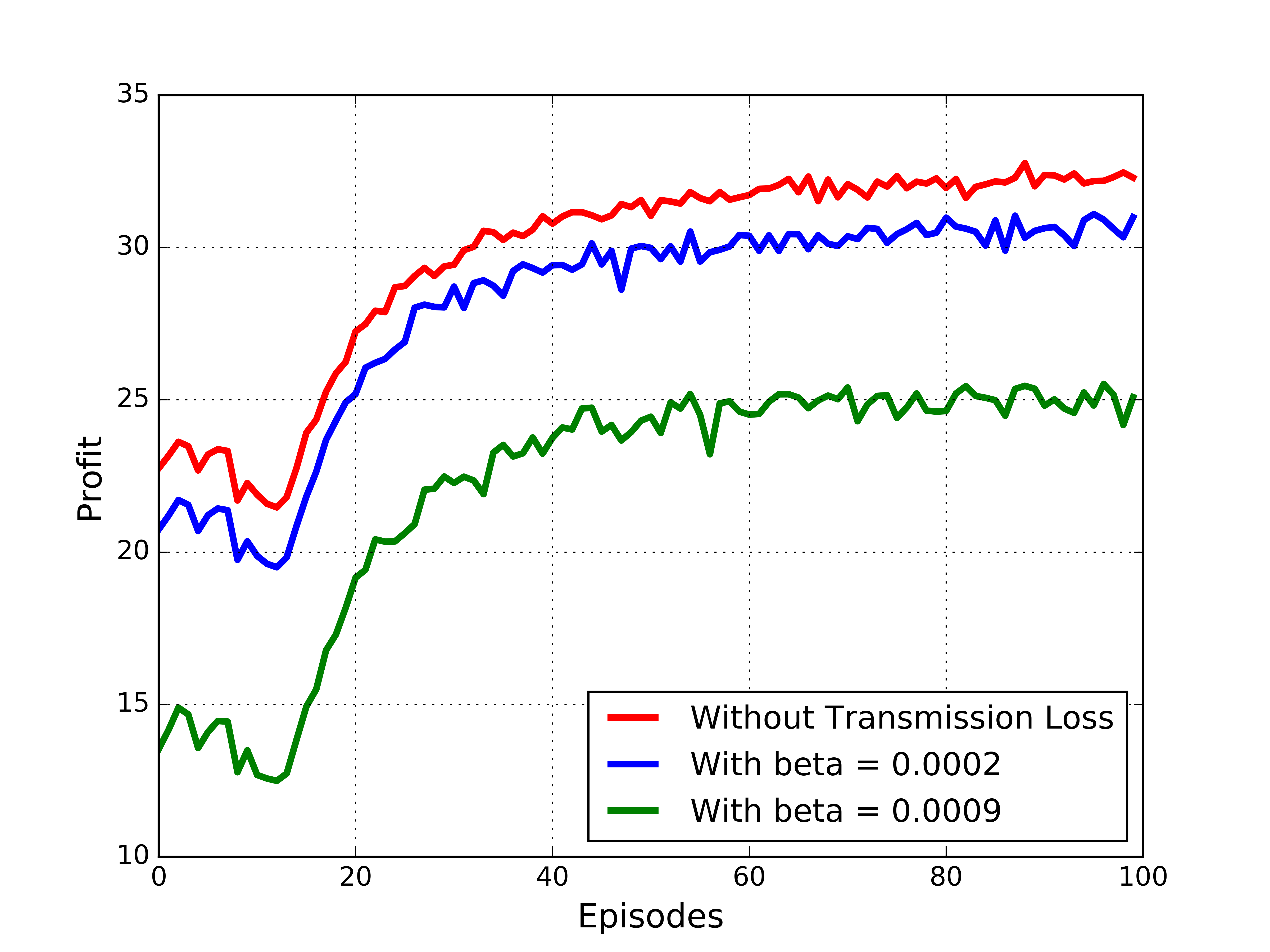}
\caption{Impact of transmission losses on total grid profit.
}
\label{Trasmission_loss}
\end{figure}

Figure \ref{Trasmission_loss} demonstrates the impact of transmission losses on the total grid profit. In particular, the results show that as the value of transmission loss $\beta$ increases, the service provider profit decreases since it has to purchase more power to meet the total demand at any given time.


\vspace{-.20cm}


\subsection{Scenario II: Medium Scale Simulation with 50 Prosumers}

\begin{figure}[t]
\includegraphics[scale=0.205, trim=2.5cm 0.7cm 0.05cm 2.2cm, clip]{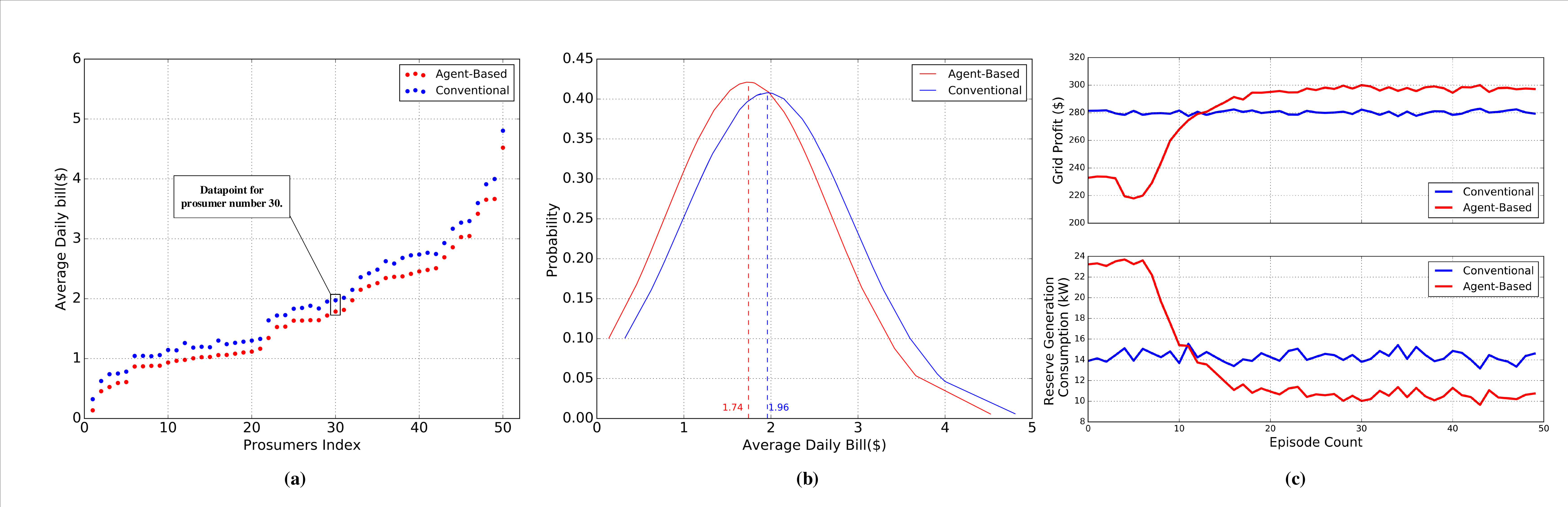}

\caption{Performance of the Conventional vs. Agent-Based scenarios in the medium scale microgrid with 50 prosumers. (a) Average daily bills of prosumers; (b) Distribution of average daily bills; (c) Performance comparison of the Conventional vs. Agent-Based scenarios in terms of grid profit and reserve power consumption with $50$ prosumers.
}
\label{50Prosumers_profit}
\end{figure}

The second scenario studies a medium scale microgrid with $50$ prosumers and $N_c = 40$ non-generational consumers.
In this scenario, the proposed method's scalability is the main criteria under investigation.
The average daily electricity bills for 50 prosumers each with a distinct consumption profile, while using Conventional and Agent-Based approaches are illustrated in Figure~\ref{50Prosumers_profit}(a). According to this figure, the average daily bill is reduced for all 50 prosumers in the Agent-Based scenario. Figure~\ref{50Prosumers_profit}(b) represents the probability distribution function of the average daily bill for 50 prosumer data points. This figure illustrates an average bill of $\$1.96$ for Conventional method, vs. $\$1.74$ for Agent-Based method, amounting to around $12\%$ bill reduction for a 24 hour billing cycle. 

\begin{figure}[t]
\centering
\includegraphics[scale=0.35, trim=0.5cm 0.1cm 0.5cm 1cm, clip]{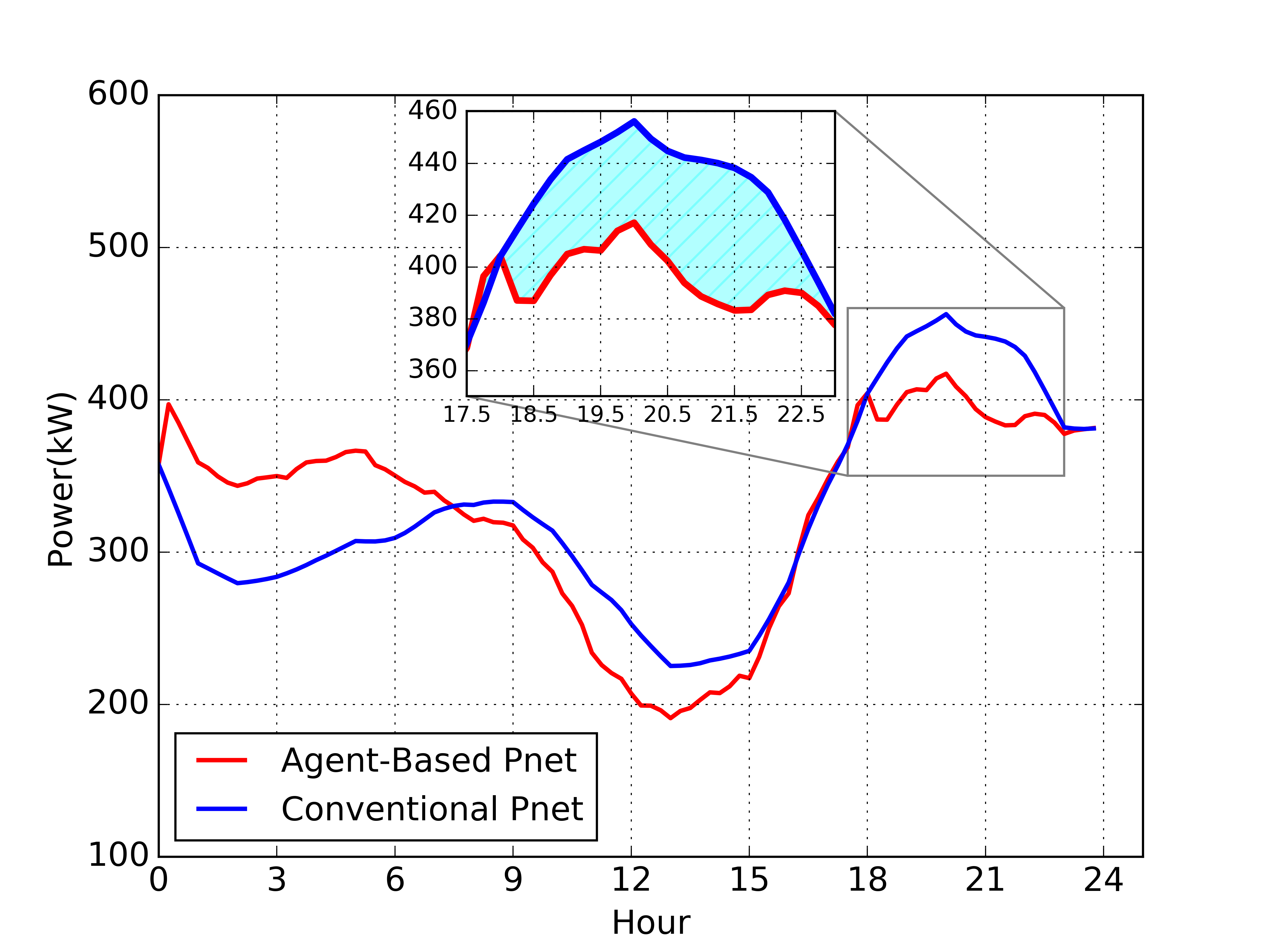}
\caption{Average net power comparison for the Conventional vs. Agent-Based scenarios during a 24-hour period. 
}
\label{AveragePnet}
\end{figure}

Figure~\ref{50Prosumers_profit} (c) compares the accumulative grid profit and reserve power utilization of the Conventional and Agent-Based scenarios in our medium scale microgrid. As pictured, the Agent-Based method provides around $4.3\%$ higher profit over a 24-hour period. This improvement is due to the fact that the SPA and PAs learn how to dispatch the batteries' power to rely on the prosumers' PV generation, rather than using the costly reserve power. 
Moreover, properly incentivizing the prosumers to dispatch their stored energy has a positive impact on shaving the peak load demand of the grid, as compared in Figure~\ref{AveragePnet} over a 24-hour period. 
As pictured, the Agent-Based approach shifts the demand at the peak hours between 18pm to 24pm to early morning hours before 6am.  

%% file: chapters/7.Conclusion.tex
\section{Conclusion}\label{sec:conclusion}
In this paper, we propose a Deep Reinforcement Learning (DRL) framework for energy management and demand response in prosumer dominated microgrids. The service provider DRL agent dynamically changes the electricity buy price and determines the direction and amount of power flow according to the household's load demand. Further, the prosumers DRL agent controls the battery charge and discharge rate and amount of power injection into the grid. These DRL agents collectively provide a dynamic decision-making framework. 
Our  simulation results demonstrate that the proposed framework provides higher economic benefits for both power grid and prosumers. Specifically, properly incentivizing prosumers through dynamic pricing and leveraging the capacity of distributed battery resources result in: (i) reduced average daily bills for prosumers, (ii) enhanced profits for the grid by decreasing the reserve generation power demand, and (iii) reduced net power demands during peak hours. 
The proposed framework can be extended by including other external observations such as weather conditions, or investigating the performance of DRL agents under peer-to-peer energy sharing  across prosumers.

